# О ВОЗМОЖНОСТИ ОБНАРУЖЕНИЯ ЭМИССИОННЫХ ЛИНИЙ НАНОАЛМАЗОВ В ОКРЕСТНОСТЯХ АСТРОФИЗИЧЕСКИХ ОБЪЕКТОВ: ДАННЫЕ ЛАБОРАТОРНОЙ СПЕКТРОСКОПИИ И НАБЛЮДЕНИЙ

**Ширяев[1,2] А.А., Сабин[2] Л., Валявин[1] Г., Галазутдинов[4] Г.**
[1]*ИФХЭ РАН, Москва, Россия*
[2]*Instituto de Astronomia, UNAM, Ensenada, Mexico*
[3]*САО РАН, Н.Архыз, Россия*
[4]*Instituto de Astronomia, Univ. Catolica del Norte, Antofagasta, Chile*
shiryaev@phyche.ac.ru; a_shiryaev@mail.ru

# ON POSSIBILITY OF DETECTION OF EMISSION FROM NANODIAMONDS IN VICINITY OF STELLAR OBJECTS: LABORATORY SPECTROSCOPY AND OBSERVATIONAL DATA

**Shiryaev[1] A.A., Sabin[2] L., Valyavin[3] G., Galazutdinov[4] G.**
[1]*IPCE RAS, Moscow, Russia*
[2]*Instituto de Astronomia, UNAM, Ensenada, Mexico*
[3]*SAO RAS, N.Arkhyz, Russia*
[4]*Instituto de Astronomia, Univ. Catolica del Norte, Antofagasta, Chile*

*Based on extensive laboratory characterization of presolar nanodiamonds extracted from meteorites, we have proposed a novel approach to detect nanodiamonds at astrophysical objects using the 7370 Å emission band arising from lattice defects. Details of laboratory spectroscopic studies and preliminary results of observations are presented.*

Since 1987, it is known that some types of meteorites contain nanodiamonds (ND). Isotopic composition of implanted noble gases indicates that some nanodiamonds should be related to supernovae explosions [1]. These nanodiamonds are characterized by log-normal size distribution from ~1 to 10 nm with median size around 2.6 nm, precluding analysis of individual grains. According to theoretical modeling and laboratory experiments, NDs are fairly stable at sizes less than few nanometers. In laboratory, NDs were synthesized by very different ways.

Attempts to observe nanodiamond features in astrophysical spectra are rather numerous, but very few of them can be (relatively) unambiguously assigned to diamonds. Most of these attempts use IR spectroscopy. As shown in [2, 3], the only real possibility of detection of nanodiamonds in IR is based on eventual observations of hot hydrogenated grains. Indeed, in [4, 5], observation of characteristic C-H features at 3.43 and 3.53 microns in emission spectra of several Herbig Ae/Be stars were reported. Perfect match of these bands to peculiar configuration of C-H bonds on surfaces of hot (800–1000K) "large" nanodiamonds (about 50 nm) makes the assignment of the observed bands to heated nanodiamonds plausible.

Spatially resolved studies [6, 7] showed that the diamond-related emission originated from the inner region (<15 AU) of the circumstellar dust disk, whereas PAH emission extended towards the outer region. Note that the "diamond" bands are observed in less than 4% of the studied Herbig stars [8].

Another approach to detect nanodiamonds is based on emission lines (photoluminescence, PL) from defects in diamond structure. The PL from nanodiamonds is usually broad and is due to surface graphitic (sp2) carbon, making the spectrum similar to G9–K0 objects and is not very characteristic. Note that, in some cases, this broad-band PL may be absent [9]. However, diamonds may contain point defects, giving rise to strong emission lines under, e.g., UV excitation. Extended Red Emission was ascribed to the photoluminescence of nitrogen-vacancy (NV) complexes in nanodiamond particles with sizes approx. 100 nm or larger [10]. However, a real stellar source emits quasi-continuum spectrum, and in this case, several charge states of the NV defects will be observed. The overlap of these contributions will make the band too broad to account for the astronomical observations (see Fig. 2 of [10]). Note that the NV defects are not observed in real nanodiamonds extracted from meteorites [2].

Importantly, spectroscopic properties of nanodiamonds are strongly size-dependent, and up to now no reliable astrophysical observations of features resembling spectra of nanodiamonds similar to those from meteorites are known.

Recently, wereported observation of an important point defect, the silicon-vacancy complex (SiV), in real nanodiamonds from meteorites [2,3]. Subsequent studies demonstrated that the SiV luminescence was confined to the smallest diamond grains with sizes below 2 nm [11]. This defect is not related to SiC grains. The SiV defect is observable in absorption and in emission as a well-defined band at 7370 Å with FWHM ~15–20 Å (Fig. 1). Strong temperature dependence of the SiV luminescence permits observation only of particles with temperatures below ~450K (the colder, the better). We suggest that observation of the SiV defect luminescence may assist in search for astrophysical sources of nanodiamonds with grain sizes comparable to those observed in meteorites. Absorption spectroscopy of this defect may complement IR observations of C-H bands pronounced in IR emission spectra of hot particles (>800–1000 K).

Combined analysis of information about structure and chemical impurities suggests that the growth process of (at least) rather abundant nitrogen-containing grains should be very fast [2]. The CVD-like (Chemical Vapour Deposition) process, possibly triggered by a shock wave(s), currently looks the most plausible one. Analysis of important details of nanodiamond synthesis in laboratory shows that finding astrophysical environment suitable for very fast growth of impurity-rich nanodiamond grains is not trivial, since it must combine relatively high density of ions (to allow condensation), rather peculiar pressure-temperature conditions (to prevent conversion to sp2-carbons) and moderate UV-flux (to prevent sublimation).

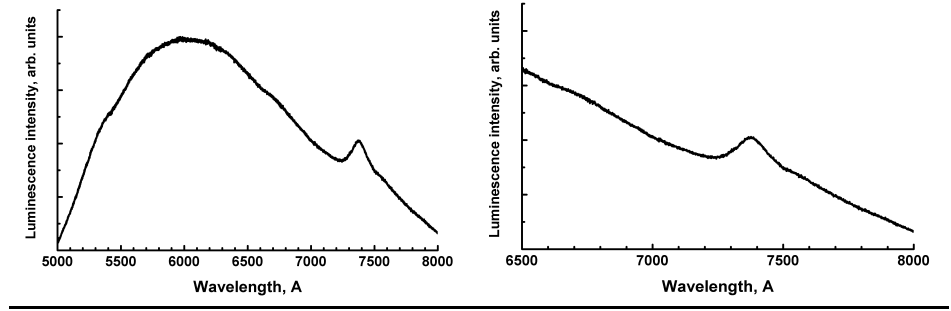

Fig.1. Laboratory spectrum of real meteoritic nanodiamonds. Left: broad spectral range, right: the SiV emission band at 7370Å.

Colliding winds from binary WR stars may be a promising site for rapid formation of nanodiamonds: dust condenses during motion along the front created by colliding stellar winds from the components. Moreover, a fraction of nanodiamonds with SN-related isotopically-anomalous xenon also fits into this scenario. In order to be implanted into nm-sized nanodiamonds, the relative speed of Xe ions should not exceed 100–200 km/sec. Such conditions could take place during interaction of the expanding shell of supernova explosion of a WR component in a binary system with the outflowing dust. Obviously, the binary WR stars are not the only promising candidates. One of significant problems in the search for the SiV emission is the absence of any reliable estimate of the signal strength, since neither nanodiamond concentration nor quantum efficiency of the SiV emission at excitation with a given stellar spectrum are known.

An observational program aiming to detect the emission line of the SiV defects from nanodiamonds around binary dust-generating Wolf-Rayet (WR104, 125, 140, 140a, 141), Herbig Ae/Be, and several carbon (HD 110914, HD 108105, AX Cyg) stars was initiated. Selected objects are promising sites for nanodiamonds formation due to reasons given above; the SiV luminescence band can be efficiently excited by broad band UV (and less efficiently by visible) radiation from the star. The observational strategy consisted of observation of the star itself and subsequent observation of the star's vicinity at distances up to 16 arcsec. Contribution of the star itself was used for proper correction of background, telluric lines, and eventual contribution of C(IV), which is close in position to the required wavelength 7370 Å. Spectral observations of WR137 and WR140 were carried out at the 6 m Russian telescope (BTA) at the Special Astrophysical Observatory (SAO) (18/19 June, 2016) using upgraded Nasmyth focus Main stellar spectrograph-polarimeter MSS (R15000, the 7000–8000 Å spectral range). The MSS is a moderate-beam classic spectrograph equipped with a circular polarization analyser combined with a 7-layer image slicer [12, 13 and http://www.sao.ru/hq/lizm/mss/en/]. Primary data reductions were made using MIDAS software packages, for details see [14]. Spectroscopic observations of other targets were performed with the 2.1 m telescope at the San Pedro Martir Ob-servatory on June 19-23, 2017. We used the Boller & Chivens spectrograph with the 13.5 μm Spectral 2 CCD (2048 × 2048 pixels) and the 1200 l/mm grating providing a dispersion of 0.60 Å/pixel and covering the 6775–7965 Å range. The 5´ long slit, 200 μm in width was, positioned on the surroundings of each WR star. A typical exposure was 1800s, and a HeNeAr lamp was obtained after each science exposure for wavelength calibration.

Unfortunately, evaluation of the data does not reveal any significant features at the required wavelength range. The strategy of observations should be modified and extensive search in available archives for high signal/noise ratio spectra is currently underway.

G.V. and G.G. acknowledge the support of the Russian Science Foundation (project 14-50-00043, area of focus: Exoplanets) for support of the experimental part of this work with data obtained at SAO RAS. We thank the staff of the San Pedro Martir Observatory for their support, in particular, I. Plauchu Frayn, F.J. Montalvo Rocha, and G. Melgoza Kennedy.